\documentstyle[12pt,aaspp4]{article}

\begin{document}

\title{Stellar Variability Background in OGLE-I Microlensing Search}

\author{Przemys\l aw Wo\'zniak$^1$ and Micha{\l } Szyma\'nski$^2$}
\affil{$^1$ Princeton University Observatory, Princeton, NJ 08544--1001, USA}
\affil{e-mail: wozniak@astro.princeton.edu}
\affil{$^2$ Warsaw University Observatory, Al. Ujazdowskie 4, 00-478 Warszawa,
Poland}
\affil{e-mail: msz@sirius.astrouw.edu.pl}

\begin{abstract}

We search OGLE-I photometric database for stars, which, as defined by formal
criteria adopted by OGLE-I microlensing search, showed variability
during only one out of 3 or 4 observing seasons.
The results include 17 previously reported microlensing events, 2 newly
discovered candidate events and 15 intrinsically variable stars that have
a potential of contaminating samples of microlensing events.
Based on photometry obtained in 1992 and 1993 OGLE \#10 was tentatively
included in the list of microlensing candidates, however its light curve
in 1994 and 1995 shows many characteristics of the variable stars
found in our search, and most likely it is not a microlensing event.
For all stars which passed our tests, we provide $44 \times 44$ arcsec
($101 \times 101$ pixels)
centered subframes from each OGLE-I frame in $I$ band. It is the first
time when images used to derive photometry of microlensing events are
available in convenient format to astronomical community.

\end{abstract}

\keywords{ Gravitational Microlensing -- Photometry -- Stars: Variables}

\section{Introduction}
The search for rare cases of gravitational microlensing in the Local Group
requires monitoring of $\sim 10^6$ stars over several months in order to
yield a significant rate of detections. A common implementation adopted
is to construct a massive photometry database and subsequently select
stars which experienced brightening of the type we expect on theoretical
grounds (see Paczy\'nski, 1996, for a recent review of basic theory,
current microlensing searches and results).
For the vast majority of events the light curves should follow
a single point mass microlensing curve. Possible departures and exceptions
from this basic case are extremely interesting. For random distribution
of stars the probability
that a given star will be lensed twice over duration of the experiment
is negligible. Large fraction of binary lenses is expected to give
raise to the population of ``wide binary'' light curves with two
separate amplification regions ($\sim 1\%$ of the total rate of events).
However in most cases the secondary peak should be weak
($A_{\rm max} < 0.1$) and probably would only be discovered in light
curves, which
called attention because of the primary event (Di Stefano and Mao 1996).
Photometry in very crowded fields, which are natural places
to look for microlensing events, is often of limited quality. The majority 
of stars measured are just above the detection limit,
many events have modest amplitudes, and numerous light curves are unevenly
sampled. Therefore a microlensing curve may accidentally give a good fit
to the brightening
which is due to the intrinsic variability of the star. Certainly a repeated
brightening of the same type would need to be very carefully examined
before any claim of a detection of the wide binary microlensing repeater.
As a result, it is generally hard to lower confusion rate
without lowering the number of events returned by the procedure.

For microlensing events reported by OGLE-I project (Udalski et al. 1992)
the basic requirement was that variability should occur during only one
observing season (Udalski et al. 1994a).
It was assumed that a sample of stars selected according to the above
condition contains the majority of microlensing events and relatively
few variable stars of other types.
In this paper we investigate how many variable stars have light curves that,
given the time sampling of OGLE-I experiment, appear constant during all
seasons except for just one. The quantitative information about stellar 
variability background, against which microlensing events are detected, 
allows further tuning of the methods used in automated detection of the events.

\section{Data}

The data used in this paper is stored in OGLE-I database of measurements
obtained during four observing seasons between 1992 and 1995.
Only fields observed for 3 or 4 seasons were analyzed. The relevant
part of the database consists of 20 fields: BW1-8, BWC, MM5-AB, MM7-AB
(4 seasons of data) and BW9-11, MM1-AB, MM3, GB1 (no data in 1992).
There were typically four months of observing time per season resulting
in $\sim 45$ frames per field per season. All observations were made using
1-m Swope telescope at the Las Campanas Observatory operated by Carnegie
Institution of Washington. A single Ford (Loral)
$2048 \times 2048$ CCD was used with the pixel scale of 0.44 arcsec/pixel,
covering $15 \times 15$ arcmin on the sky.
Images in the database were de-biased and flat-fielded
by an automated data pipe line based on IRAF ``ccdred'' routines.
Photometry was performed using DoPhot package (Schechter, Mateo and Saha 1995).
We refer to Udalski et. al (1992) for details of the photometric data pipeline
and to Szyma\'nski and Udalski (1993) for a description of the database. After
preliminary selection based on photometry alone, we also inspected actual
images for 94 objects (see Section~3). From each good quality
frame in $I$ band OGLE-I database (grades A--E, F frames rejected),
we extracted a $101 \times 101$ pix ($44 \times 44$ arcsec) subframe centered
on the star of interest.
One of the main goals of this project is to provide such set of subframes
for all stars which are described in our paper. This is the first
data set of this type; actual frames used to derive photometry of
microlensing events are available for a variety of studies, including
independent photometry obtained with entirely different software
(e.g. Alard and Lupton 1998).
At the end of this article we provide an ftp address.

\section{Search criteria}

An estimate of the optical depth to microlensing requires full understanding
of the selection effects. Udalski et al. (1994a) designed an automated
selection procedure for microlensing events in order to facilitate
calculation of the sensitivity function. A basic requirement for
candidate stars was that they should vary during one observing season
and stay constant during remaining seasons.
Then more stringent conditions followed, i.e., a  star should increase
its brightness rather than decrease and a standard microlensing curve
should yield a significant improvement of the fit with respect to a constant.
Here we apply only the first group of conditions and we put no
restrictions on variability type other than a genuine change in brightness
during one of the observing seasons.

First we require at least 75 ``good'' photometric measurements in $I$ band
for stars observed during 3 seasons and 100 such measurements
for stars with 4 seasons of data. This prevents many potential
false alarms due to spurious stars which are inevitably detected by
DoPhot, and does not noticeably affect the final sample,
as the great majority of the objects are usually well measured.
A ``good'' measurement comes from a frame of grade A--D (E, F rejected),
has stellar DoPhot type and a standard deviation returned by DoPhot
is less than 1.6 times median of DoPhot standard deviations for the
entire set of measurements of this particular star.

\noindent
A star is considered variable during a given season when:

\begin{enumerate}
\item
   At least 5 consecutive ``good'' points deviate consistently up or down
   from the average over ``constant'' seasons by more than
   $3 \sigma_{\rm max}$,
   where $\sigma_{\rm max}$ is a function of the stellar magnitude and field
   (Udalski et al. 1993).

   or

\item
   There are at least 10 such points total.

\end{enumerate}

\noindent
A star is considered constant during remaining 2 or 3 seasons if:

\begin{enumerate}
\item
   There are at least 40 ``good'' measurements in $I$ band after
   $\pm 4$ ``raw'' $\sigma$ rejection.

\item
   Dispersion of the ``good'' measurements is less than $\sigma_{\rm max}$
   for a star of a given magnitude in a given field.

\item
   The star's mean magnitude from all ``good'' photometric points is $I<19.5$.

\end{enumerate}

Stars fainter than $V=19.0$ were also tested for presence of close companions,
as in this case seeing variations introduce large uncertainties in their
photometry. The minimum allowed distance for the nearest star was a linear
function of its magnitude. We used empirical formula of Udalski et al. (1993):
$R_{crit}=0.875 \times (21.0-I)$ pixels.

The above criteria were programmed and filters were run on the $I$ band
light curves from OGLE-I database. Typically 60 stars per field were
selected, $\sim 1200$ stars total out of $\sim 2 \times 10^6$. There are
two previously reported microlensing events which did not pass this test.
We discuss those exceptions in the next section.

Despite our efforts, it proved impossible to check reality of the recorded
magnitude variations for stars which passed the test without visual
inspection of the images. This is due to relatively large
variety of CCD defects that were present in various parts of the
frame and also bleeding columns whose impact on a particular star changes
from frame to frame.
We also could not afford extracting relevant pixels from each of the OGLE-I
frames for 1200 stars. Therefore, we inspected light curves of the
selected objects by eye
and rejected the ones which showed purely erratic changes. Some of the
stars had light variations correlated in time and position in the frames,
a clear indication of a problem, usually bleeding column.
Such objects were also removed.
We left 94 objects, ``promising'' ones and also some material for case studies
of the anticipated sources of spurious variability.
For those stars we extracted
a $101 \times 101$ pixels subframe from each frame in the OGLE-I database.
These subframes were then checked for any possible factors which may have
affected the photometry. Typical bad measurements were caused by
physical CCD defects or bleeding columns.
In some cases seeing variations in extremely crowded environment
influence photometry despite good statistics.
We found that of all object in the OGLE-I database which passed the above
tests, 35 experienced a real change in brightness
(OGLE \#2a and \#2b counted as two).

\section{Results}
\subsection{Microlensing events}

As we start the search with a subset of the conditions used by OGLE-I
experiment to extract microlensing candidates, we should, in principle,
recover all events previously found in the automated search by Udalski
et al. (1994a). We do indeed find all events used
for the estimate of the optical depth to microlensing in the direction
of the Galactic Bulge. All remaining events reported by OGLE-I so far
were detected in real time by the Early Warning System
(hereafter EWS; Udalski et al. 1994b), except for OGLE \#8, which was found by
non-algorithmic means in some preliminary analysis of the BW9 field,
but could not be included in the optical depth calculation.
Observations of BW9 field started in 1993 and at the time Udalski et al.
(1994a)
published their results the data for only one observing season were available.
It turns out that OGLE \#8 is blended with a relatively bright star,
close enough not to satisfy the condition for acceptable blends (Section~3).
Moreover OGLE \#19 due to its large amplitude was caught by
the EWS, although it is fainter than $I=19.5$.
For completeness we include those two events in the final data set.

OGLE \#13 is the first event detected on-line by MACHO collaboration
(Alcock et al. 1994) that was followed up by the OGLE group
(Szyma\'nski et al. 1994).
It is located in GB5 field, which we do not analyze here.
OGLE \#10, most likely a variable star, is discussed in the next section.

Two of the newly discovered objects have light curves which are best
interpreted as magnification due to microlensing. These are new OGLE-I
microlensing candidates, OGLE \#20 and OGLE \#21. They were missed in the
automated search by Udalski et al. (1994a), who had to compromise the number
of detected events in order to eliminate human judgment.

In Table~1 we summarize basic information on OGLE-I microlensing events:
identification numbers in $I$ and $V$ databases for a given field, position,
source magnitude in unlensed state $I_{\rm s}$, $V-I$ color,
fitted maximum amplification $A_{\rm max}$, time scale $t_0$ (in days)
and moment of the maximum light $t_{\rm max}$ (Heliocentric $\rm JD -
2448000$). Please note that $t_0$ used here is the time it takes the
source to move by one Einstein radius in the sky relative to the
lensing mass, while MACHO group is using duration of the event, $2 t_0$.
We assume no blending, i.e., that the amplified source accounts for 100\% of
the light measured within the PSF long before and long after the event.
The observed
shift of light centroid as the lensed star brightens (Goldberg and Wo\'zniak
1998) and estimates based on artificial frames by Goldberg (1998) suggest
that a fraction of strongly blended events should be large.
The standard fit is irrelevant
in case of OGLE \#7 -- a binary event (Udalski et al. 1994c).
Figure~1 shows full light curves
along with the corresponding fit from Table~1. $30 \times 30$ arcsec
finding charts are shown in Figure~4. A microlensed star is always at the
center, as indicated by a white cross, north is up and east is to the left.
Table~3 may be used to identify stars in the database of $V$ measurements.
A number starting with ``N'' indicates that the object was not detected
in the template image and its photometry is stored in the database
of the so called ``new'' objects (Szyma\'nski and Udalski 1993).

\subsection{Variable stars}

We found 15 variables which were formally constant during all seasons
except for one. Full light curves are presented in Figure~2.
It is quite surprising that most of them seem to show some regularities,
taking into account that the set of conditions described in Section~3
strongly discriminates against periodicity.
To quantify this first impression, we looked for periods between
0.1 and 100 days in those light curves using the Analysis of Variance method
and following the approach of Udalski et al. (1994d).
Table~2 contains results of this analysis.
For all 15 stars we list: identification number in $I$ database
for a given field, position, mean $I$ magnitude and mean $V-I$ color
from the entire set of ``good'' measurements along with the amplitude
$\Delta I$.
We detected periodicity in light curves of 12 stars (top portion of
Table~2). For those objects we give the detected period $P$ (in days)
and the moment of the maximum brightness $T_0$ ($\rm JD ~Hel. - 2448000$)
chosen to fall into the observing season, in which the star was found to be
variable.
Figure~3 shows corresponding phased light curves.
An even more unexpected result is that four of the stars which passed
the test were already found by the OGLE team in searches for strictly
periodic variables ! (Udalski et al. 1995a, 1995b, 1996, 1997).
The last column of Table~2
contains previous OGLE identifications, where appropriate.
We do not attempt any further analysis of these objects and confine
ourselves to the statement that they closely resemble miscellaneous
variables described in detail by Olech (1996), majority of which
probably have spotted surfaces.

The last 3 stars in Table~2 have no significant period in the range
0.1--100 days.
Stars BW10 $I$ 179313 and BW10 $I$ 184744 may still have very long periods,
while MM3 $I$ 58214 went through a single outburst. In this case $T_0$ is the
moment of the maximum measured brightness, while $P$ is the time interval
between $T_0$ and the moment when the star reached half of its maximum
brightness. Finding charts for all objects from Table~2 are given in Figure~4.

OGLE \#10 was tentatively reported as a candidate microlensing event by
Udalski et al. (1994a). It came out of the automated procedure and at the
time of discovery the data showed brightening during 1992 observing season
and flat light curve during season of 1993. Nevertheless in 1994 and 1995
the star developed semi regular variations with the period around 36 days
superimposed on overall brightening with a much longer time scale. The first
feature is characteristic for variability of the majority of the stars
classified as miscellaneous in the OGLE-I catalog (Olech 1996), and the second
is reminiscent of the star BW10 $I$ 179313.
Therefore most likely OGLE \#10 is a variable star similar to the ones
analyzed by Olech (1996), of which quite many examples we find in our search.

\section{Concluding remarks.}

A general conclusion is that variability background in OGLE-I search was
reasonably well separated from microlensing events, although OGLE \#10,
a single candidate which most likely is not a microlensing event,
constitutes a 5\% confusion rate. A variable star, just like any other star,
may be amplified by microlensing, but an increase of brightness by 0.1 mag
may be naturally explained by intrinsic variability of this star, especially
that similar objects (certainly not microlensing events) apparently exist.

We find two additional possible events that were not returned by the automated
procedure of Udalski et al. (1994a). This is not surprising since we relax some
of the selection cuts applied before. Moreover, one of those events happened
near the end of the observing season while the other had very short time
scale and poorly sampled light curve. Both of them are very inconspicuous.

The outburst experienced by MM3 $I$ 58214 (most likely a flare or CV star)
is an important case which
has a potential of contaminating samples of microlensing candidates.
Suppose we had no data just before the flare. With photometric accuracy
comparable to OGLE-I data such variable could be taken for a fading
microlensing event and uncertainty would have to be resolved
by spectroscopy and/or monitoring of the star long after the event.

We note a relatively large number of periodic or almost periodic
variables which change amplitude. They mimic constant stars for time long
enough to pass the most important criterion of the OGLE-I search,
i.e., that a star should vary within a limited time interval with essentially
constant flux at all other times. It is mostly due to relatively poor
time coverage of the OGLE-I experiment and should not be difficult to overcome
in the second phase of the project, OGLE-II (Udalski, Kubiak and Szyma\'nski
1997).
Two recommendations can be made for the future. First, low amplitude
events, e.g. with $A_{\rm max} < 1.5$, may be safely ignored in
the calculation of the optical depth to prevent potential problems
with objects similar to OGLE \#10. MACHO team is already using
such cut off to avoid contamination by ``bumpers''. Second,
a requirement of roughly even photometric coverage of both sides of the
magnification peak allows filtering out stars with (usually
asymmetric) outbursts. OGLE-I events used in the optical depth
determination satisfy such condition, however some of the
events discovered by the EWS do not.

The full set of data used in this paper, including $101 \times 101$ pix
subframes extracted from every $I$ band image of each object in
Tables~1 and 2, is available for public. Images in FITS format
as well as photometric data in standard Johnson system may be retrieved
via anonymous ftp from {\tt astro.princeton.edu (128.112.24.45)} -- directory
{\tt /ogle/var\_background} and {\tt sirius.astrouw.edu.pl (148.81.8.1)}
-- directory /ogle/var\_background. See {\tt README} file for details.

\acknowledgments{We would like to thank Prof. Bohdan Paczy\'nski for
encouragement and helpful suggestions during this project.
Comments from David Goldberg allowed us to improve the manuscript.
This work was supported with NSF grant AST--9530478.
MSz was partly supported by KBN grant BST to Warsaw University Observatory.}

\newpage


\hoffset=-1.5cm

\begin{deluxetable}{lrrcccccccc}
\tablewidth{0pt}
\tablecaption{OGLE-I microlensing events.}
\tablehead{
\colhead{Field} &
\colhead{Star ID} &
\colhead{Star ID} &
\colhead{$\alpha_{2000}$} &
\colhead{$\delta_{2000}$} &
\colhead{$I_{\rm s}$} &
\colhead{$V-I$} &
\colhead{$A_{\rm max}$} & \colhead{$t_0$} & \colhead{$t_{\rm max}$} & \colhead{OGLE} \\
\colhead{} & \colhead{$I$} & \colhead{$V$} &\colhead{}&\colhead{}&
\colhead{mag} &
\colhead{}&\colhead{}&\colhead{ d } & \colhead{[JD]\tablenotemark{*}} & \colhead{\#} }
\startdata
  BW7 & 117281 & 110374 & 18:04:24.80 & -30:05:58.3 & 18.77 & 1.52 & 2.35 &  21.3 & 1154.2 &  1 \nl
%
%
  BW5 & 178651 & 239408 & 18:02:52.08 & -30:04:21.7 & 19.20 & 1.72 & 6.34 &  46.3 &  804.6 & 2a \nl
  BWC &  10648 &   9055 & 18:02:52.02 & -30:04:20.2 & 19.18 & 1.66 & 5.12 &  50.3 &  804.7 & 2b \nl
  BW3 & 161225 &  93508 & 18:04:43.45 & -30:14:10.7 & 15.91 & 1.80 & 1.25 &  11.5 &  830.7 &  3 \nl
  BW4 & 111538 &  N8860 & 18:04:16.27 & -29:51:56.6 & 19.29 & 1.65 & 4.41 &  14.0 &  806.9 &  4 \nl
  BWC & 120698 &  97013 & 18:03:21.84 & -30:02:32.0 & 18.01 & 1.55 & 7.36 &  13.3 &  824.3 &  5 \nl
MM5-B & 128727 &  69133 & 17:47:45.19 & -35:01:18.4 & 18.16 & 1.45 & 3.08 &   8.6 &  818.8 &  6 \nl
%
%
  BW8 & 198503 & 189096 & 18:03:35.83 & -29:42:00.8 & 17.53 & 1.80 &  ... &   ... &    ... &  7 \nl
  BW9 & 138910 & 125157 & 18:00:49.20 & -29:47:05.6 & 17.91 & 2.06 & 2.09 &  45.6 & 1218.1 &  8 \nl
MM7-A &  86776 & N22929 & 18:10:41.90 & -25:50:33.9 & 19.22 & 1.69 & 1.93 &  16.6 &  814.7 &  9 \nl
  BW6 & 167045 & 217426 & 18:03:45.13 & -30:18:16.9 & 18.20 & 1.74 & 1.31 &  12.0 & 1537.4 & 11 \nl
  BW5 &  83758 & 112206 & 18:02:24.62 & -30:07:51.3 & 18.70 & 2.02 & 2.00 &  18.3 & 1582.6 & 12 \nl
MM1-A & 123474 & 104845 & 18:06:48.67 & -26:37:24.2 & 19.05 & 2.35 & 1.81 &  11.4 & 1822.1 & 14 \nl
  BW3 & 142477 &  97949 & 18:04:37.26 & -30:12:11.5 & 18.35 & 1.76 & 3.97 &  15.6 & 1853.1 & 15 \nl
  BW5 &  72182 &  94317 & 18:02:07.60 & -30:01:12.3 & 18.43 & 1.56 & 1.84 &  26.2 & 1884.9 & 16 \nl
 BW10 & 176006 & 164907 & 18:00:56.60 & -29:58:17.6 & 18.80 & 1.92 & 2.00 & 121.9 & 1999.2 & 17 \nl
  BW1 &  67895 &  49873 & 18:02:11.10 & -29:51:21.7 & 18.63 & 1.55 & 2.25 &   9.0 & 1918.1 & 18 \nl
  MM3 &   3289 & N90940 & 18:07:28.53 & -29:11:30.9 & 19.77 & 1.15 & 6.62 &   5.0 & 1931.2 & 19 \nl
MM5-A &  41271 & N34210 & 17:46:57.68 & -34:47:14.4 & 19.15 & 1.74 & 5.23 &  30.0 &  858.6 & 20 \nl
  BW6 &  80736 & 104108 & 18:03:10.79 & -30:15:34.2 & 18.81 & 1.92 & 2.02 &   5.1 & 1162.1 & 21 \nl
\enddata
\tablenotetext{*}{Heliocentric $\rm JD - 2448000$}
\end{deluxetable}

\begin{deluxetable}{lrrcccccccc}
\tablewidth{0pt}
\tablecaption{Variable Stars found in the search.}
\tablehead{
\colhead{Field} &
\colhead{Star ID} &
\colhead{Star ID} &
\colhead{$\alpha_{2000}$} &
\colhead{$\delta_{2000}$} &
\colhead{$\left<I\right>$} &
\colhead{$V-I$} &
\colhead{$\Delta I$} & \colhead{$P$} & \colhead{$T_0$} & \colhead{OGLE} \\
\colhead{} & \colhead{$I$} & \colhead{$V$} &\colhead{}&\colhead{}&
\colhead{mag} &
\colhead{}&\colhead{}&\colhead{ d } & \colhead{[JD]\tablenotemark{*}} & \colhead{ID} }
\startdata
  BW2 & 205380 & 166732 & 18:02:48.18 & -30:18:23.2 & 15.92 & 1.98 & 0.14 & 37.59 &  730.90 & V28 \nl
  BW3 & 127604 &  88469 & 18:04:36.95 & -30:21:10.5 & 15.85 & 2.14 & 0.13 & 26.26 &  725.70 &     \nl
  BW3 & 161220 & 108232 & 18:04:46.43 & -30:14:10.7 & 15.77 & 2.03 & 0.14 & 35.98 & 1824.82 &\#10 \nl
  BW5 &  10705 &  13570 & 18:01:54.55 & -30:03:21.1 & 14.68 & 1.99 & 0.11 & 79.76 & 1552.19 & V10 \nl
  BW9 & 160297 & 146622 & 18:00:59.81 & -29:53:05.9 & 15.49 & 2.23 & 0.16 & 24.11 & 1830.25 & V21 \nl
 BW11 &  74701 &  64198 & 18:00:41.46 & -30:18:27.4 & 15.84 & 2.21 & 0.18 & 24.88 & 1145.92 &     \nl
MM1-B & 198992 & 194344 & 18:07:08.19 & -26:48:52.0 & 17.24 & 2.20 & 0.17 & 84.96 & 1823.35 &     \nl
  MM3 &  30724 &  31147 & 18:07:39.85 & -29:10:38.3 & 16.95 & 2.19 & 0.18 & 19.35 & 1823.88 &     \nl
  MM3 &  89081 &  90520 & 18:07:41.33 & -29:01:10.8 & 15.64 & 2.13 & 0.13 & 43.86 & 1531.39 &     \nl
  MM3 & 227272 & 236470 & 18:08:19.38 & -28:57:25.9 & 15.70 & 1.84 & 0.15 & 15.72 & 1135.94 &     \nl
MM5-A &  17258 &   6871 & 17:46:51.07 & -34:44:03.1 & 15.55 & 2.32 & 0.13 & 60.87 & 1547.82 & V17 \nl
MM5-A & 199110 &  89237 & 17:47:56.95 & -34:51:34.7 & 14.23 & 1.99 & 0.19 & 33.96 & 1823.67 &     \nl

&&&&&&&&\nl
 BW10 & 184744 & 172256 & 18:01:08.64 & -30:07:10.7 & 17.04 & 2.42 & 0.21 &   ... &     ... &     \nl
 BW10 & 179313 & 167019 & 18:00:55.29 & -29:56:21.3 & 15.45 & 2.05 & 0.10 &   ... &     ... &     \nl
  MM3 &  58214 &  56998 & 18:07:33.83 & -29:00:37.8 & 19.49 & 1.27 & 1.20 &  4.52 & 1176.82 &     \nl
\enddata
\tablenotetext{*}{Heliocentric $\rm JD -2448000$}
\end{deluxetable}

\newpage
\pagestyle{empty}
\hoffset=0cm

\begin{figure}[t]
\figurenum{1a}
\plotfiddle{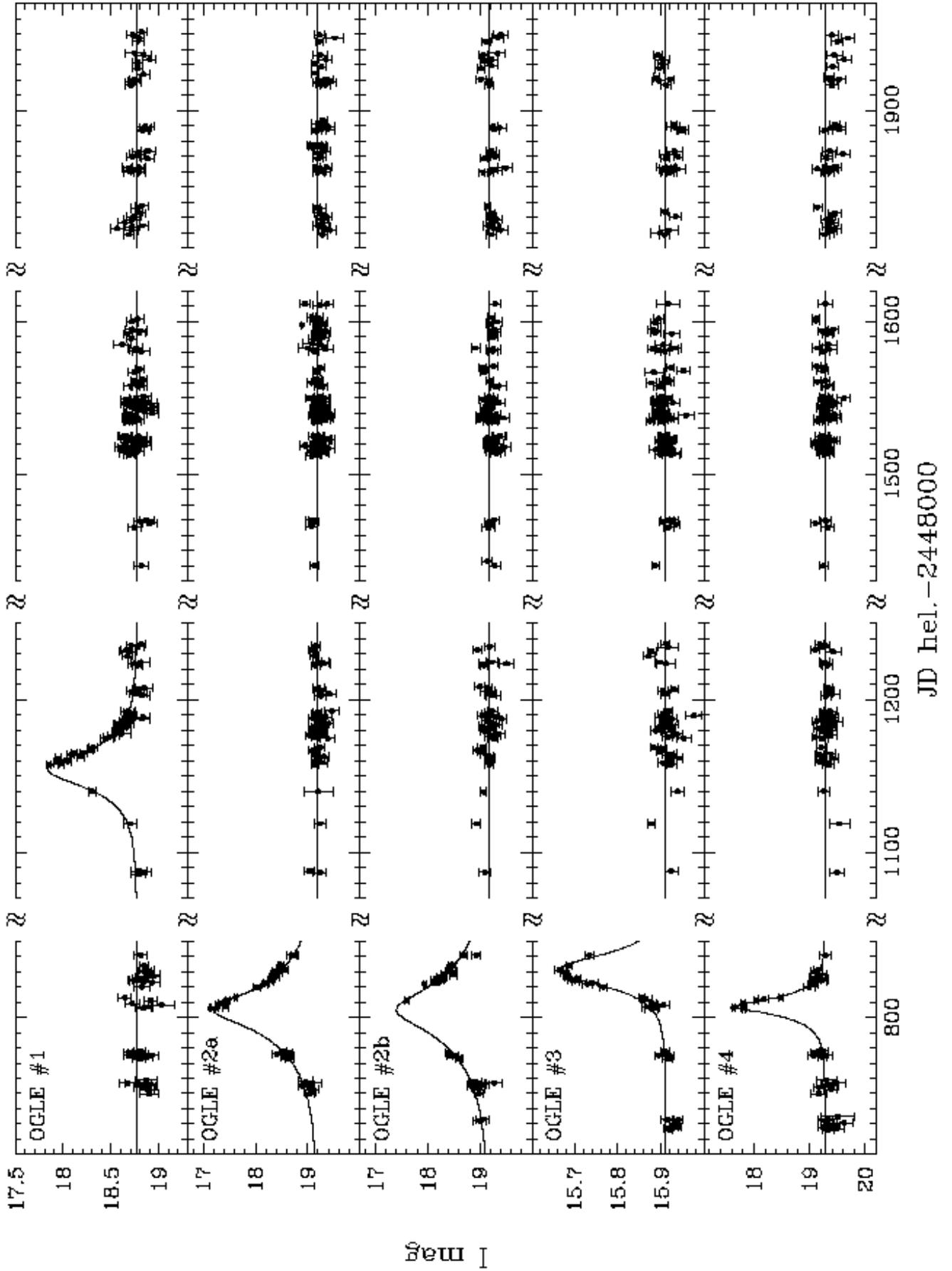}{21cm}{0}{100}{100}{-300}{-60}
\caption{OGLE-I microlensing events.}
\label{fig:lens1}
\end{figure}

\begin{figure}[t]
\figurenum{1b}
\plotfiddle{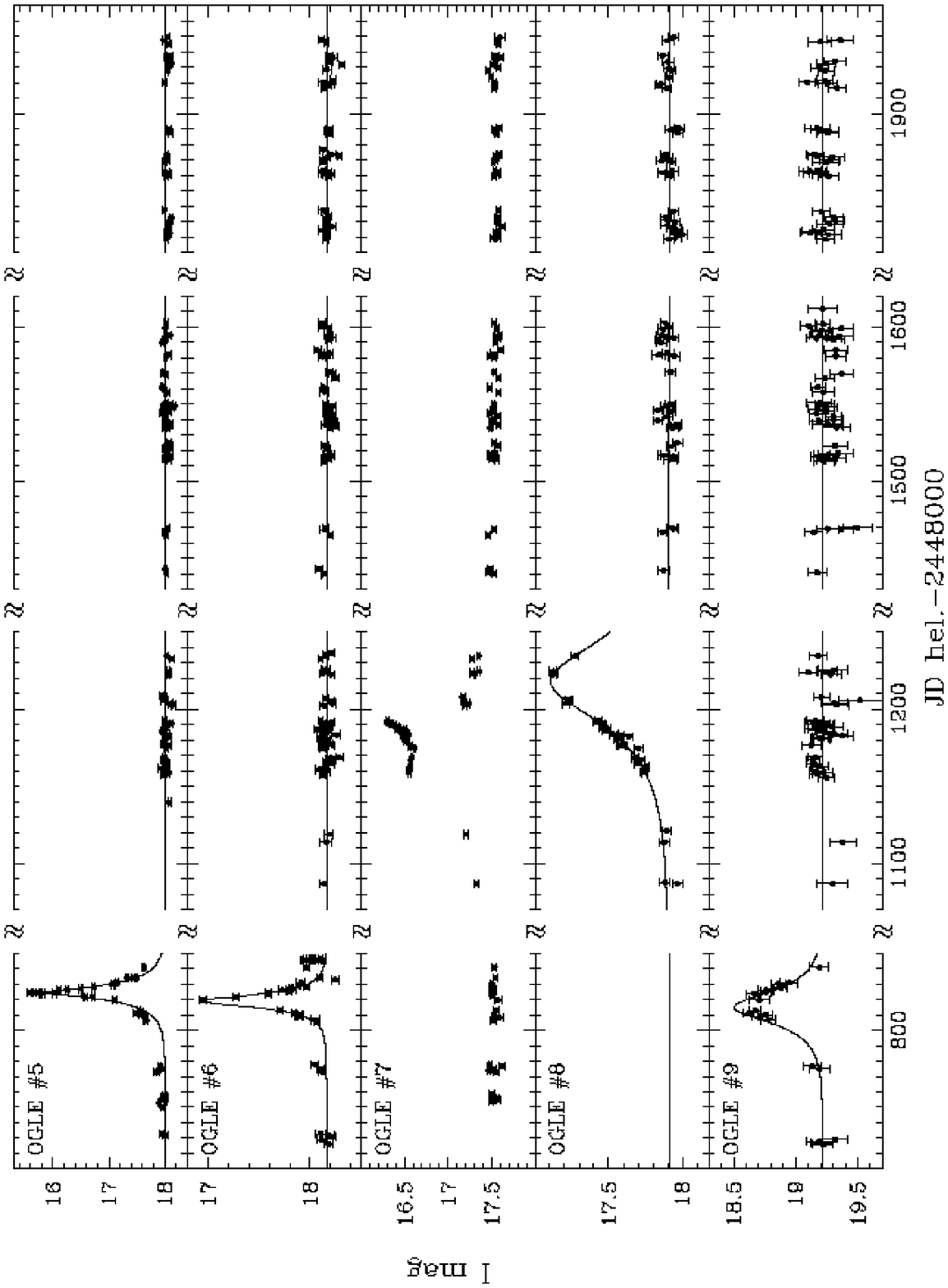}{21cm}{0}{100}{100}{-300}{-60}
\caption{OGLE-I microlensing events -- continued.}
\label{fig:lens2}
\end{figure}

\begin{figure}[t]
\figurenum{1c}
\plotfiddle{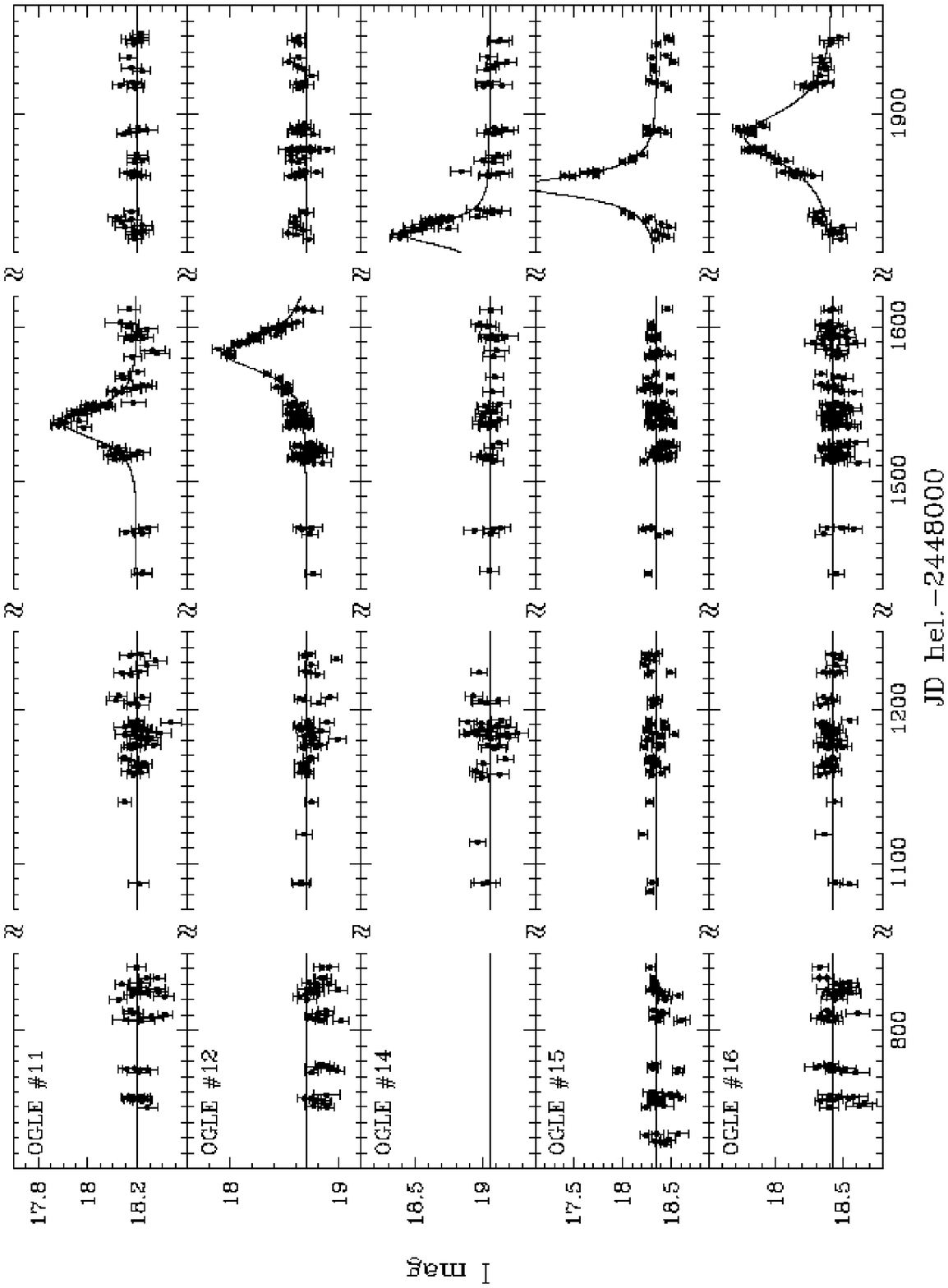}{21cm}{0}{100}{100}{-300}{-60}
\caption{OGLE-I microlensing events -- continued.}
\label{fig:lens3}
\end{figure}

\begin{figure}[t]
\figurenum{1d}
\plotfiddle{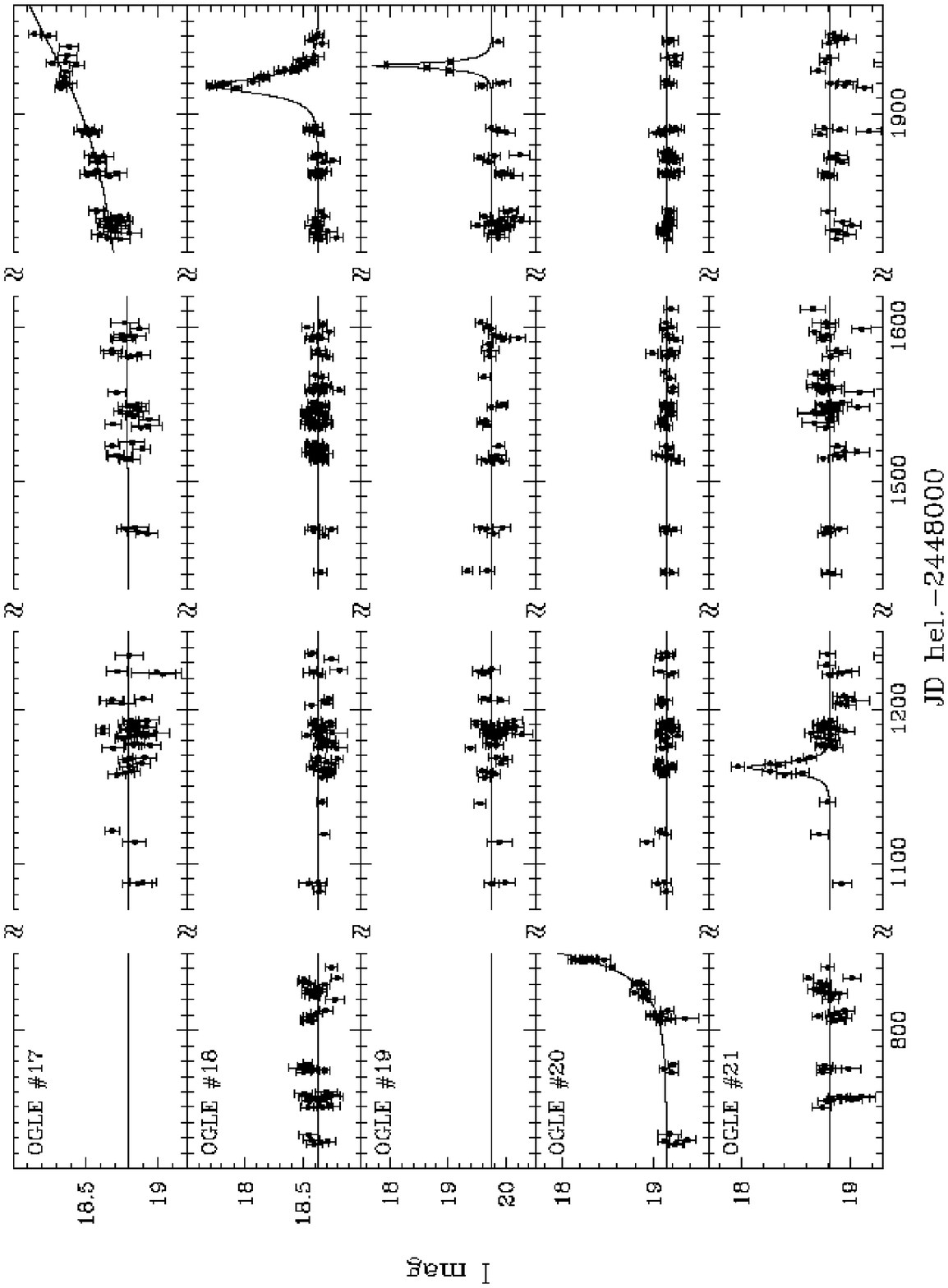}{21cm}{0}{100}{100}{-300}{-60}
\caption{OGLE-I microlensing events -- continued.}
\label{fig:lens4}
\end{figure}

\begin{figure}[t]
\figurenum{2a}
\plotfiddle{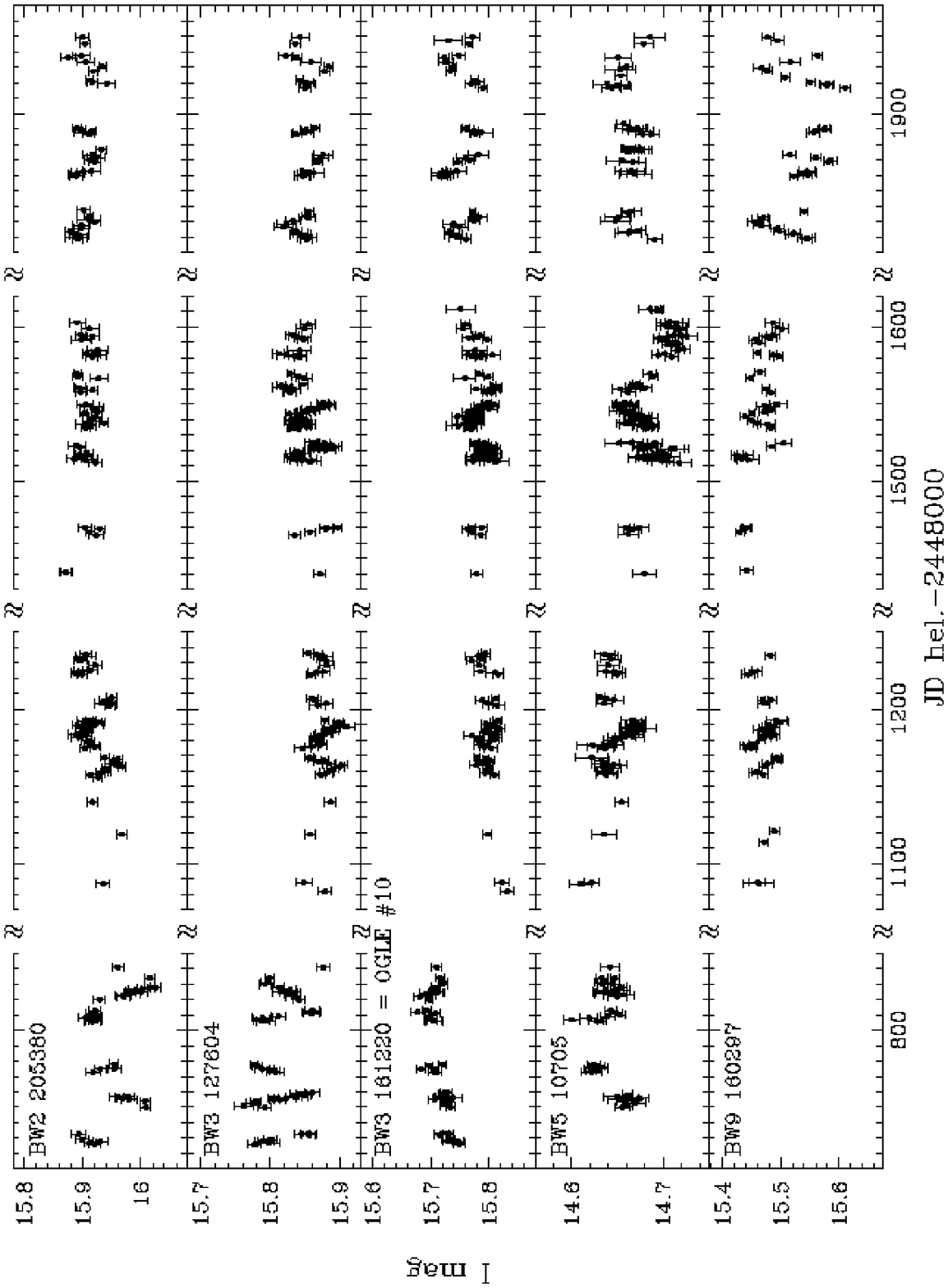}{21cm}{0}{100}{100}{-300}{-60}
\caption{Variable stars found in the search.}
\label{fig:var1}
\end{figure}

\begin{figure}[t]
\figurenum{2b}
\plotfiddle{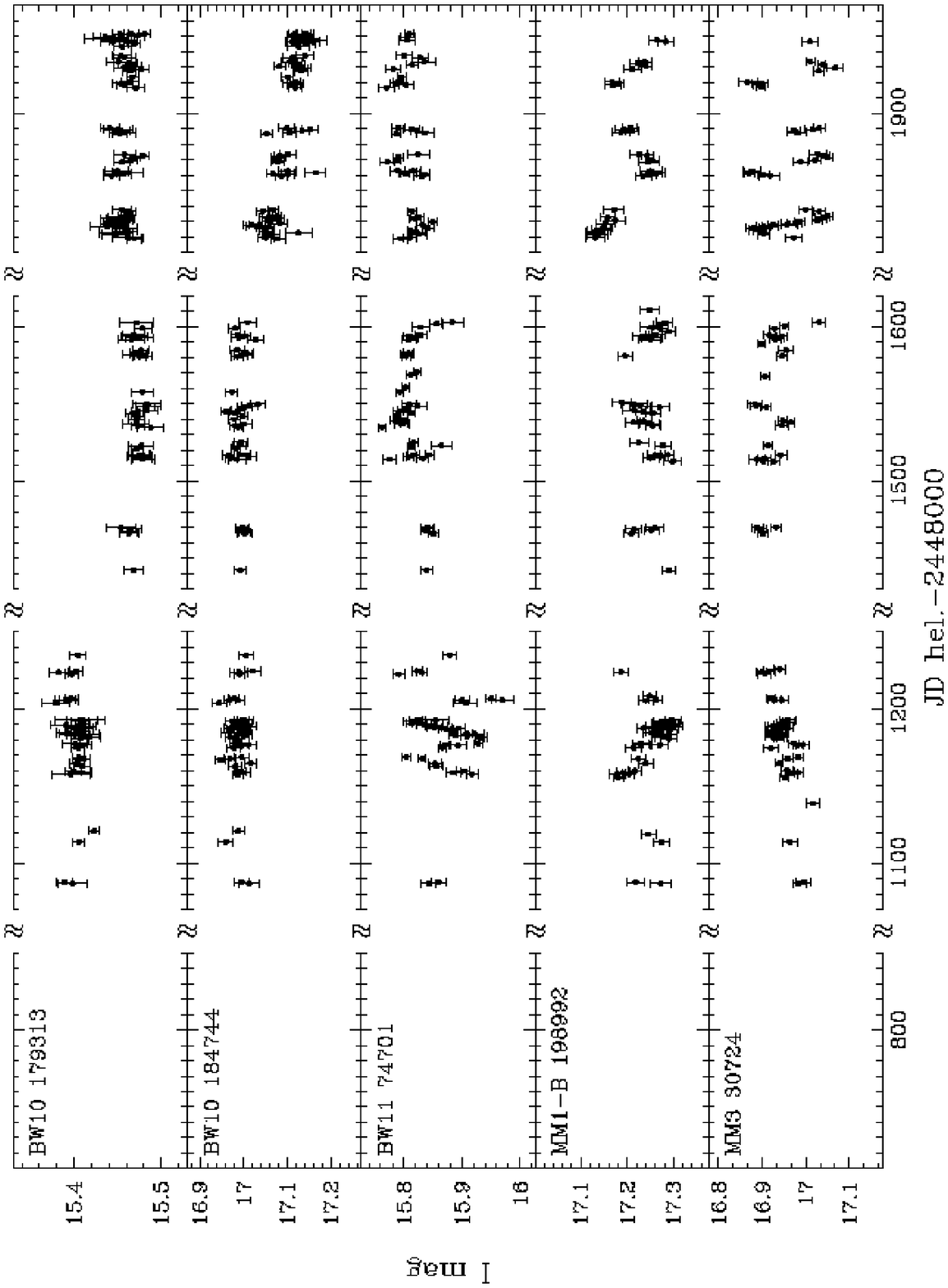}{21cm}{0}{100}{100}{-300}{-60}
\caption{Variable stars found in the search -- continued.}
\label{fig:var2}
\end{figure}

\begin{figure}[t]
\figurenum{2c}
\plotfiddle{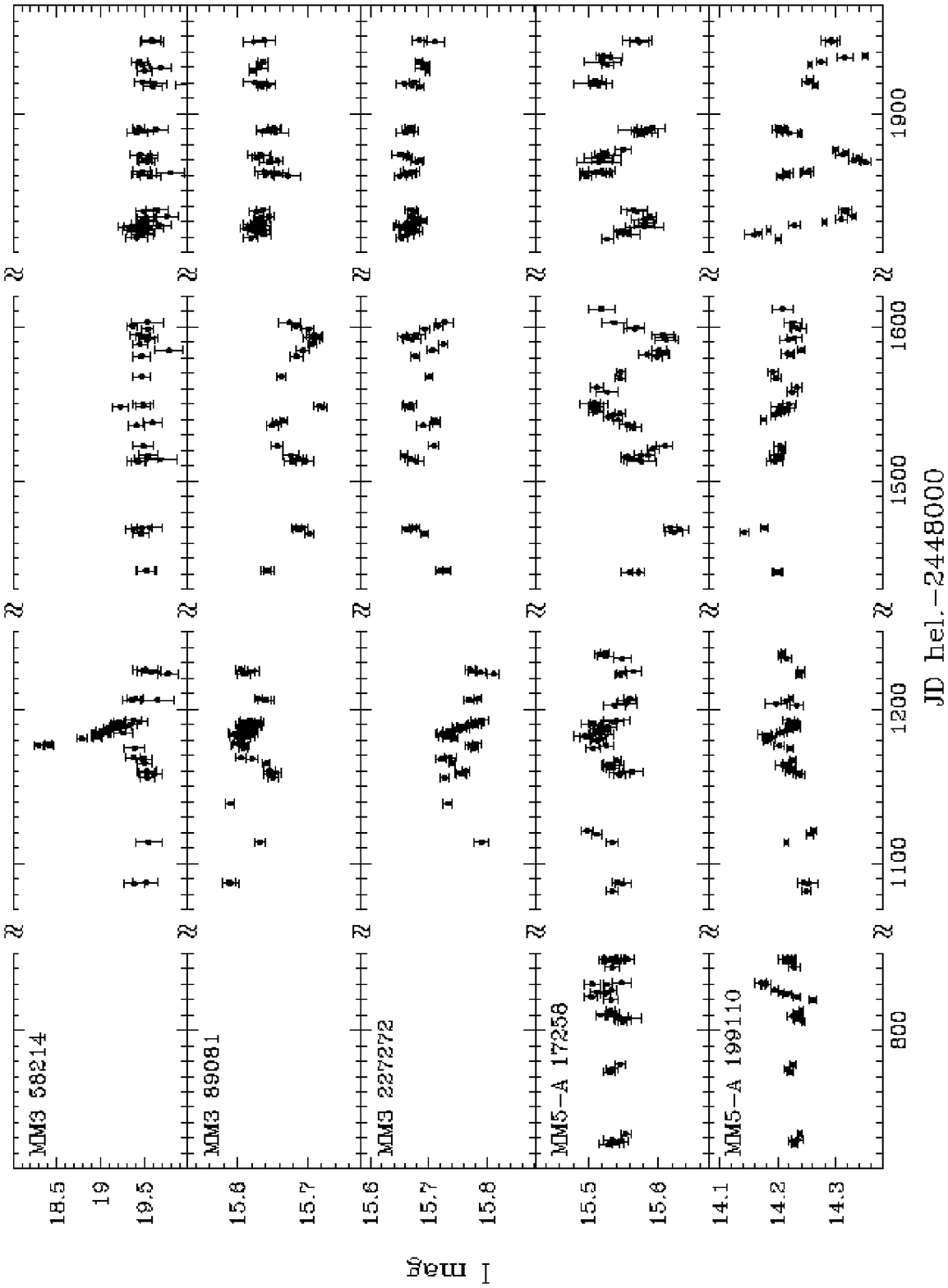}{21cm}{0}{100}{100}{-300}{-60}
\caption{Variable stars found in the search -- continued.}
\label{fig:var3}
\end{figure}

\begin{figure}[t]
\figurenum{3}
\plotfiddle{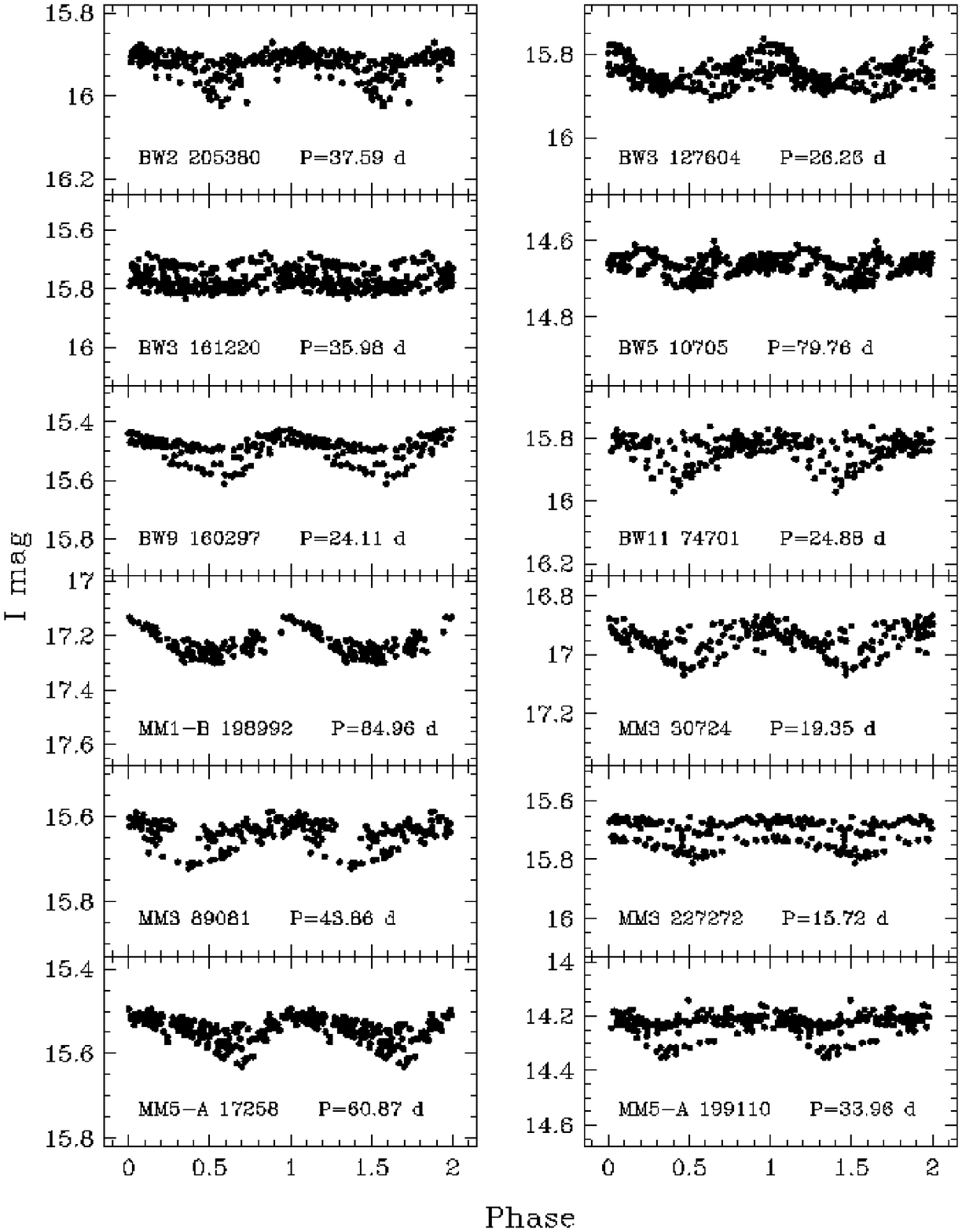}{21cm}{0}{100}{100}{-300}{-60}
\caption{Results of the test for periodicity.}
\label{fig:phased}
\end{figure}

\begin{figure}[t]
\figurenum{4a}
\caption{Finding charts.}
\label{fig:fcharts1}
\end{figure}

\begin{figure}[t]
\figurenum{4b}
\caption{Finding charts -- continued.}
\label{fig:fcharts2}
\end{figure}

\end{document}